# V$_2$C MXene–modified g-C$_3$N$_4$ for enhanced visible-light photocatalytic activity

Ruizheng Xu, Guiyu Wei, Zhemin Xie, Sijie Diao, Jianfeng Wen, Tao Tang,

Li Jiang, Ming Li [a,*], Guanghui Hu [a,*]

College of Science & Key Laboratory of Low-dimensional Structural Physics and Application, Education Department of Guangxi Zhuang Autonomous Region, Guilin University of Technology, Guilin 541004, China

[*]E-mail addresses: 2006017@glut.edu.cn (M. Li), guanghui@glut.edu.cn (G. Hu).

ABSTRACT

Increasing the efficiency of charge transfer and separation efficiency of photogenerated carriers are still the main challenges in the field of semiconductor-based photocatalysts. Herein, we synthesized g-$C_3N_4$@$V_2C$ MXene photocatalyst by modifying g-$C_3N_4$ using $V_2C$ MXene. The prepared photocatalyst exhibited outstanding photocatalytic performance under visible light. The degradation efficiency of methyl orange by g-$C_3N_4$@$V_2C$ MXene photocatalyst was as high as 94.5%, which is 1.56 times higher than that by g-$C_3N_4$. This was attributed to the $V_2C$ MXene inhibiting the rapid recombination of photogenerated carriers and facilitating rapid transfer of photogenerated electrons ($e^-$) from g-$C_3N_4$ to MXene. Moreover, g-$C_3N_4$@$V_2C$ MXene photocatalyst showed good cycling stability. The photocatalytic performance was higher than 85% after three cycles. Experiments to capture free radicals revealed that superoxide radicals ($\cdot O_2^-$) are the main contributors to the photocatalytic activity. Thus, the proposed g-$C_3N_4$@$V_2C$ MXene photocatalyst is a promising visible-light catalyst.

# 1 Introduction

Various synthetic dyes are directly released into water bodies, causing severe water pollution. Anionic azo dyes are the most commonly used synthetic dyes. Methyl orange (MO) is a common anionic azo dye that is very poisonous and carcinogenic to humans. Semiconductor-based photocatalysts are one of the most effective means of harvesting solar energy for tackling the energy and environmental crisis and are highly promising for reducing such water pollution issues [1]. Several semiconductor photocatalysts, including CdS, $TiO_2$, $Bi_2WO_6$, $BiVO_4$, and g-$C_3N_4$, have been studied over the last decade [2-5]. Among them, g-$C_3N_4$ is frequently utilized for pollutant degradation because of its superior performance, nontoxicity, nonmetallic composition, and distinctive electronic band structure. However, its photocatalytic activity is constrained by low light absorption, slow charge transfer, and high carrier recombination rate. Therefore, Zhuang et al. have focused on improving the separation efficiency and charge transfer rate of electron–hole pairs in g-$C_3N_4$ [6].

Modification of g-$C_3N_4$ using cocatalysts is known to improve the above properties of g-$C_3N_4$. Cocatalysts usually include various materials, such as nanoparticles, carbon nanotubes, graphene quantum dots, MXene, 2D $MoS_2$/graphene (MG), $LaFeO_3$ and $LaNiO_3$ [7-14]. For instance, the reaction rate constant k value of 2D CN/MG ternary photocatalyst prepared by in situ adsorption method was 4.8 times higher than that of pure g-$C_3N_4$ [12]. Zhang et al. synthesized Ag/g-$C_3N_4$/$LaFeO_3$ Z-scheme heterojunction photocatalysts with degradation rates of 98.97% and 94.94% for methylene blue and tetracycline hydrochloride [13]. In addition, Bao et al. fabricated highly efficient $LaNiO_3$/g-$C_3N_4$/$MoS_2$ photocatalyst, which effectively promoted water decomposition and pollutant degradation [14]. Among them, MXene, which can be used as cocatalyst in place of precious metals, have been the focus of extensive research in the last decade in the fields of environmental pollution and catalysis. $Ti_3C_2$, the first reported MXene, is widely employed in photocatalysis owing to its unique accordion-like nanosheet structure, excellent electrical conductivity, high carrier mobility, and various active groups at the surface terminus. Notably, $V_2C$ MXene has a morphology similar to that of $Ti_3C_2$ MXene but possesses a greater specific surface area and more active centers, thus making it a promising cocatalyst for pollutant degradation [15]. Zhou et al. have recently synthesized a three-dimensional porous ZnO/$V_2C$ MXene using $V_2C$ MXene as a cocatalyst [16]. $V_2C$ MXene promoted carrier separation and migration and improved the photocatalytic performance of methylene blue. A $V_2C$/BVO composite catalyst prepared through a hydrothermal method can accelerate electron transfer and optimize hydrogen evolution reaction activity [17]. In addition, Pt/$V_2CTx$ prepared by Wang et al. exhibited excellent activity and stability [18]. The above reports demonstrate that the study of $V_2C$ MXene as a cocatalyst is promising for the development of semiconductor photocatalysts. However, the degradation of pollutants using $V_2C$ MXene–modified g-$C_3N_4$ has not been previously investigated. Therefore, we explored the photocatalytic activity of g-$C_3N_4$@$V_2C$ MXene photocatalyst and applied it to the photocatalytic degradation of MO.

Herein, g-$C_3N_4$@$V_2C$ MXene photocatalyst was synthesized using a simple self-assembly mixing technique and tested for their photocatalytic degradation efficiency toward MO under visible light. The g-$C_3N_4$@$V_2C$ MXene photocatalyst exhibited higher charge transfer rates and electron–hole pair separation efficiencies than g-$C_3N_4$, considerably promoting the photocatalytic degradation of MO. A possible mechanism of the photocatalytic degradation of MO by g-$C_3N_4$@$V_2C$ MXene photocatalyst was proposed based on the analysis of relevant experimental data. The study demonstrated that $V_2C$ MXene is a cocatalyst with excellent performance and high potential for remedying water pollution.

**2 Material and methods**

2.1 Materials

Carbamide ($H_2NCONH_2$ ≥ 99%), tetramethylammonium hydroxide (TMAOH ≥ 25 wt%), and MO ($C_{14}H_{14}O_3N_3SNa$) were purchased from Aladdin. Hydrofluoric acid (HF) and ethanol ($C_2H_5OH$) were purchased from Xilong Chemical. MAX ($V_2AlC$) was purchased from Foshan Xinxi Technology Co., Ltd. Deionized (DI) water was prepared in a laboratory.

2.2 Catalyst synthesis

g-$C_3N_4$ was produced through a thermal polycondensation reaction with urea following a previously reported procedure [19]. First, 20 g of urea was dropped into a covered alumina crucible, which was then heated in a muffle furnace to 550°C at a heating rate of 2.5°C/min and maintained for 3 h at 550°C. After cooling to 25°C, the yellow solid formed in the crucible was ground into a fine powder.

$V_2AlC$ (2 g) was added to 40 mL of 49% HF and mixed well to obtain a solution. The obtained solution was stirred for 24 h at 25°C and then for 48 h at 50°C to remove the aluminum layer. Then, 40 mL of 25-wt% TMAOH was added to the solution and stirred continuously for 24 h [20-22]. The agitated solution was centrifuged several times at 8000 rpm with ethanol and DI water until the pH approached neutral; it was then freeze-dried for 24 h to obtain $V_2C$ MXene.

The self-assembly mixing technique was employed to synthesize g-$C_3N_4$@$V_2C$ MXene photocatalyst. DI water (50 mL) was combined with 0.2 g of g-$C_3N_4$ and sonicated for 30 min; then, different masses of $V_2C$ MXene were added to g-$C_3N_4$ and sonicated for 30 min. Finally, the homogeneous dispersion was obtained through magnetic stirring for 3 h [23]. The powder samples in the mixture were rinsed with DI water multiple times and dried for 24 h in a freeze-drying oven to obtain g-$C_3N_4$@$V_2C$ MXene photocatalyst(referred to as g-$C_3N_4$@$V_2C$-X, X = 10, 20, and 30 mg). Fig. 1 shows the flowchart for the production of the g-$C_3N_4$@$V_2C$ MXene photocatalyst.

2.3 Characterization

The crystal structures of the photocatalysts were analyzed through X-ray diffraction (XRD, MiniFlex-600 from JEOL, Japan). The Fourier-transform infrared (FTIR) spectra of the photocatalysts were obtained using the Nicolet-6700 infrared spectrometer (Tokyo, Japan). The morphology and elemental contents of the prepared samples were examined using scanning electron microscopy (SEM, Czech TESCAN MIRA LMS) and X-ray energy-dispersive spectroscopy. The surfaces of the samples were analyzed via X-ray photoelectron spectroscopy (XPS, Thermo Scientific K-Alpha, USA). The lattice spacing of the samples was determined through high-resolution transmission electron microscopy (HRTEM, JEOL JEM 2100Plus, Tokyo, Japan). The photoluminescence (PL) spectra were obtained using FL/FS900 (Edinburgh, Cheadle). The ultraviolet–visible (UV–Vis) absorption spectra were measured using a PerkinElmer Lambda 750 spectrophotometer.

2.4 Electrochemical measurements

A standard three-electrode electrochemical workstation (CHI 660E) was used to measure transient photocurrent and electrochemical impedance. The electrolyte was $Na_2SO_4$ (0.2 M), and the light source was a xenon lamp (PLS-SXE300, 300 W). A platinum-sheet electrode served as the counter electrode, the standard calomel electrode (SCE) served as the reference electrode, and the sample served as the working electrode. To enable the deposition of the photocatalyst onto a fluorine-doped tin oxide (FTO) electrode to form a yellowish coating, the samples were dispersed in ethanol, water, and naphthalene sulfone to obtain a mixed solution. The mixed solution was then applied onto the FTO electrode using a pipette gun. Finally, the resultant FTO electrode was dried in an oven.

2.5 Photocatalytic degradation

The photocatalytic performance and stability of each sample were investigated using a xenon lamp (PLS-SXE300, 300 W) as a visible light source and an MO solution (10 mg/L) as a contaminant. In general, 50 mg of the photocatalyst was mixed well with 50 mL of the MO solution and stirred for 30 min in the dark to attain the adsorption–desorption equilibrium of MO molecules on the photocatalyst. Afterward, the solution was placed under the xenon lamp and 3.5 mL of the irradiated solution was taken at 0.5 h intervals. After the experiment, the suspended catalyst was filtered using a 220 nm polyethersulfone needle filter. UV–Vis spectroscopy was performed to detect numerical changes in the concentration of the target dyes.

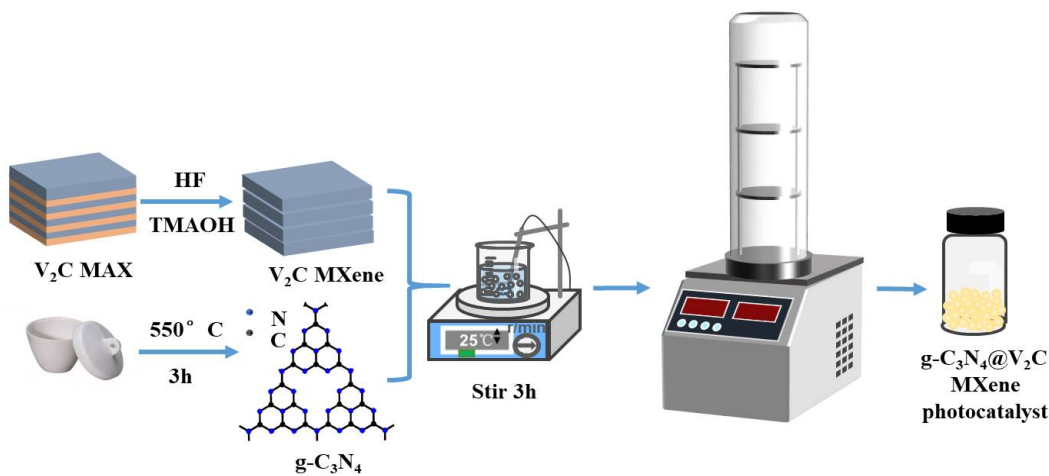

**Fig. 1.** Preparation of g-C$_3$N$_4$@V$_2$C MXene photocatalyst.

## 3 Results and discussion

3.1 Morphological and structural analyses

Fig. 2 shows the SEM images of g-C$_3$N$_4$, V$_2$C, and g-C$_3$N$_4$@V$_2$C-20 mg as well as elemental maps for g-C$_3$N$_4$@V$_2$C-20 mg. The structure of g-C$_3$N$_4$ (Fig. 2a) is layered and agglomerated, which is typical for g-C$_3$N$_4$ prepared via thermal polycondensation [24]. Pure V$_2$C MXene (Fig. 2b) exhibits a two-dimensional nanosheet-like structure. The lamellar structure of the aggregates does not substantially change after the addition of V$_2$C, proving that V$_2$C MXene exhibits a negligible effect on the basic structure of g-C$_3$N$_4$. Moreover, the distributions of C, N, and V are fairly uniform in g-C$_3$N$_4$@V$_2$C-20 mg (Figs. 2e–h). These results confirm the successful preparation of the g-C$_3$N$_4$@V$_2$C MXene photocatalyst.

For further analysis, the prepared g-C$_3$N$_4$@V$_2$C MXene photocatalyst were observed using HRTEM. Fig. 3 shows the HRTEM images of pure g-C$_3$N$_4$ and g-C$_3$N$_4$@V$_2$C-20 mg. The g-C$_3$N$_4$ presented in Fig. 3a comprises multilayered nanosheets—a typical two-dimensional layered structure [25]. In particular, because of the special structure of g-C$_3$N$_4$, no considerable lattice stripes are observed in Fig. 3a. The HRTEM image of g-C$_3$N$_4$@V$_2$C-20 mg (Fig. 3b) shows that g-C$_3$N$_4$ is in close contact with V$_2$C MXene [26]. Moreover, the V$_2$C MXene has three orientations with the same crystal spacing ($d$ = 0.15 nm). This crystal plane spacing belongs to the (011) crystal plane of V$_2$C MXene. Thus, HRTEM results verify that V$_2$C MXene successfully modified g-C$_3$N$_4$.

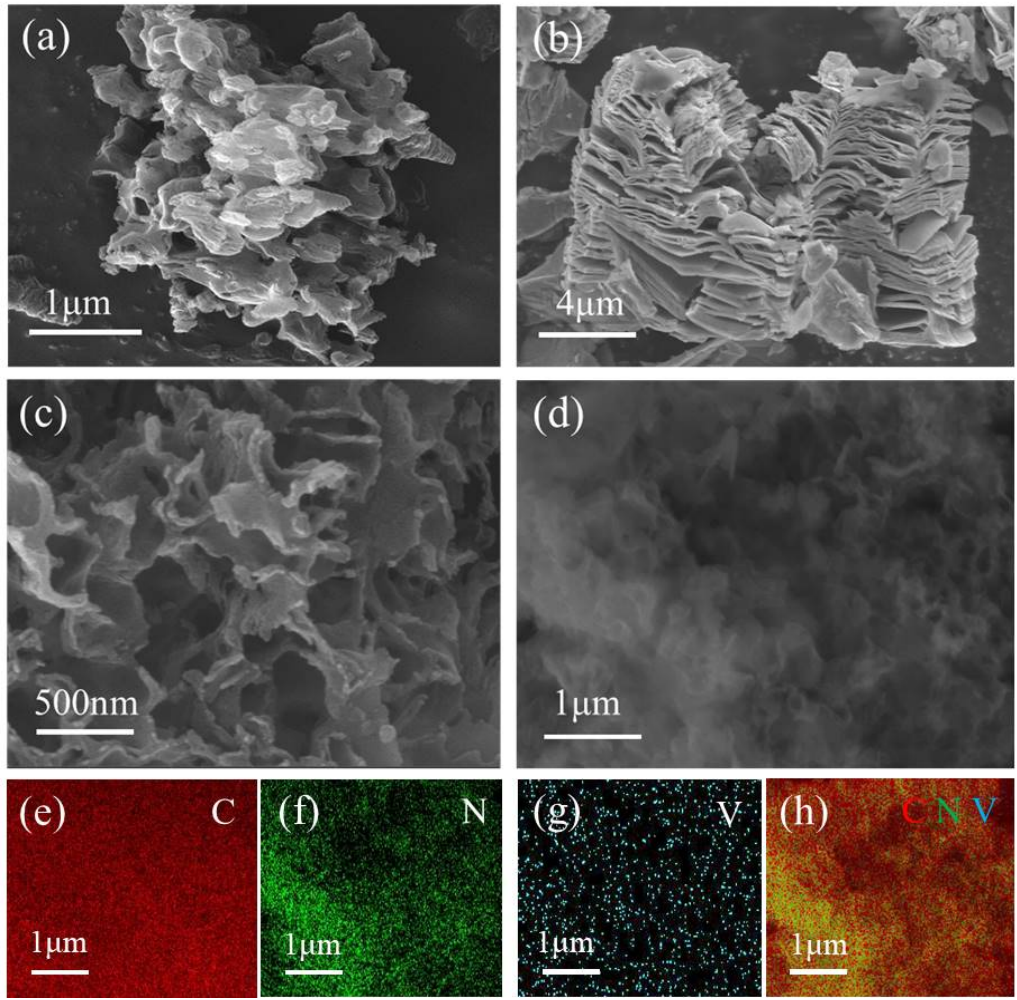

**Fig. 2.** SEM images of (a) g-$C_3N_4$, (b) $V_2C$ MXene, and (c and d) g-$C_3N_4$@$V_2C$-20 mg. (e–h) Elemental maps of the C, N, and V in g-$C_3N_4$@$V_2C$-20 mg in selected regions of (d).

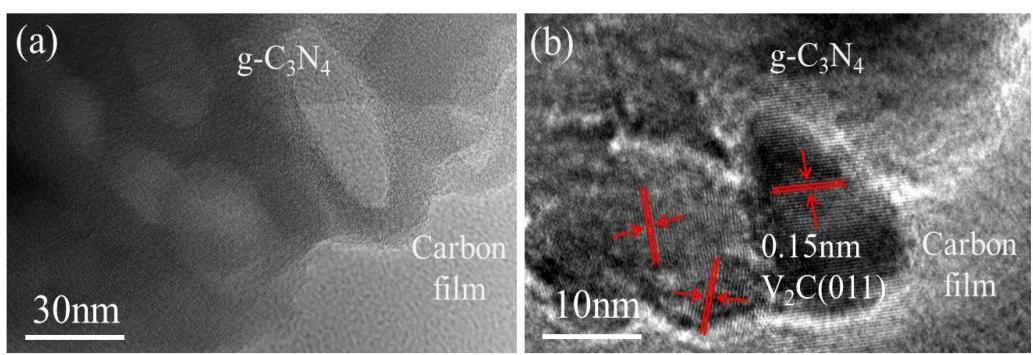

**Fig. 3.** HRTEM images of (a) g-$C_3N_4$ and (b) g-$C_3N_4$@$V_2C$-20 mg.

The crystal structures of g-$C_3N_4$, $V_2C$ MXene, and g-$C_3N_4$@$V_2C$-20 mg were determined using XRD (Fig. 4a). Both g-$C_3N_4$ and g-$C_3N_4$@$V_2C$-20 mg show two strong peaks at 13.1° (100) and 27.3° (002), which were ascribed to the interlayer reflection of the tri-s-triazine unit and the interlayer stacking of the aromatic system, respectively [27]. Interestingly, the g-$C_3N_4$@$V_2C$ MXene photocatalyst exhibits higher diffraction peak intensities and higher crystallinity compared to g-$C_3N_4$.

Compared to those of V$_2$AlC, the peaks of V$_2$C MXene are present at a lower angle and a new peak appears at 6.1°, confirming the successful preparation of V$_2$C MXene [28, 29]. Furthermore, the (010) crystal plane of V$_2$C MXene is present in the XRD pattern of g-C$_3$N$_4$@V$_2$C-20 mg. All peaks of pure g-C$_3$N$_4$ and V$_2$C MXene can be observed in the XRD pattern of g-C$_3$N$_4$@V$_2$C-20 mg, confirming the preparation of g-C$_3$N$_4$@V$_2$C MXene photocatalyst. The XRD pattern of g-C$_3$N$_4$@V$_2$C-20 mg (Fig. 4c) is essentially unaltered after cycling, and some distinctive peaks are also quite visible, indicating that the materials are very stable.

The surface groups of V$_2$C MXene, g-C$_3$N$_4$, and g-C$_3$N$_4$@V$_2$C-20 mg were analyzed using FTIR (Fig. 4d). Both pure g-C$_3$N$_4$ and g-C$_3$N$_4$@V$_2$C-20 mg exhibit three identical characteristic absorption peaks at 2970~3410 cm$^{-1}$, 1205~1690 cm$^{-1}$, and ~815 cm$^{-1}$, which were ascribed to the vibrational absorption of the NH$_X$ group, the conjugated CN heterocycle, and the tri-s-triazine unit, respectively [30]. No additional V$_2$C MXene characteristic peaks are identified in the XRD pattern of g-C$_3$N$_4$@V$_2$C-20 mg except at ~2350 cm$^{-1}$, which was attributed to the low mass fraction of V$_2$C MXene and the overlap of these peaks at 1600 and 760 cm$^{-1}$ with the vibration/stretching mode of g-C$_3$N$_4$. In summary, the characteristic absorption peaks of g-C$_3$N$_4$ and g-C$_3$N$_4$@V$_2$C-20 mg are comparable, indicating that the introduction of a moderate quantity of V$_2$C MXene does not affect the surface groups of g-C$_3$N$_4$.

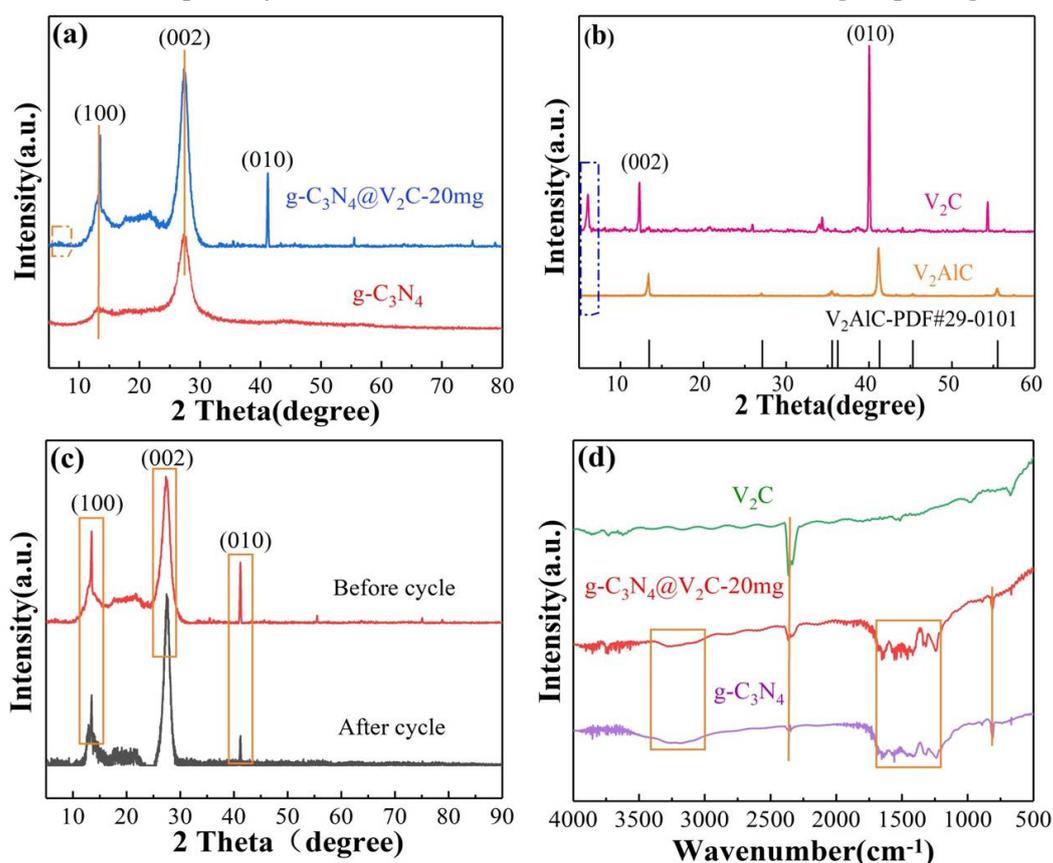

**Fig. 4.** XRD patterns of (a) g-C$_3$N$_4$ and g-C$_3$N$_4$@V$_2$C-20 mg, (b) V$_2$C MXene and V$_2$AlC MAX, and (c) g-C$_3$N$_4$@V$_2$C-20 mg (before and after cycling). (d) FTIR spectra of V$_2$C MXene, g-C$_3$N$_4$, and g-C$_3$N$_4$@V$_2$C-20 mg.

The chemical composition of the photocatalysts was determined through XPS.

The XPS spectra of the g-$C_3N_4$@$V_2C$ MXene photocatalyst (Fig. 5a) shows distinct C, N, and O peaks, where the O peak is probably due to $H_2O$/oxidation-generated oxides adsorbed on the sample surface. The absence of a V characteristic peak in the XPS spectrum of g-$C_3N_4$@$V_2C$-20 mg was attributed to the low amount of $V_2C$, resulting in an inconsequential V peak. The three major peaks in the C1s spectrum (Fig. 5b) at 284.8 (C=C), 286.5 (C–$NH_X$), and 288.2 eV (N=C-$N_2$) on the C 1s spectrum are ascribed to the aromatic ring, $sp^3$-coordinated carbon bond, and $sp^2$-hybridized carbon atom, respectively. The four binding energy peaks in the N1s spectrum (Fig. 5c) at 398.5 (C–N=C), 399.9 (N–$C_3$), 401.0 (C–$NH_2$), and 404.6 eV were ascribed to $sp^2$-hybridized N, graphitic N, amino, and excitation, respectively [31]. The five peaks shown in Fig. 5d correspond to different oxidation states of V. The $V^{5+}$ peak at 517.5 eV has the second-highest intensity. The two peaks in the V2p spectrum at 520.6 and 516.4 eV were attributed to $V^{4+}$. Finally, two $V^{3+}$ peaks are observed at 514.9 and 513.6 eV [32].

Compared with those of g-$C_3N_4$, some peaks of g-$C_3N_4$@$V_2C$-20 mg are displaced. Considerable displacement (dashed portion) of the peaks implies that $V_2C$ MXene and g-$C_3N_4$ have a strong surface interaction. Compared to those in g-$C_3N_4$, the C1s and N1s in the composite migrate to higher energy levels, indicating that g-$C_3N_4$ is an electron donor and $V_2C$ MXene is an electron acceptor [33]. Thus, $V_2C$ MXene exhibits excellent electron trapping ability. The XPS results demonstrate that $V_2C$ MXene is an efficient cocatalyst and can promote the rapid movement of photogenerated electrons from g-$C_3N_4$ to $V_2C$ MXene and enhance the visible-light photocatalytic performance.

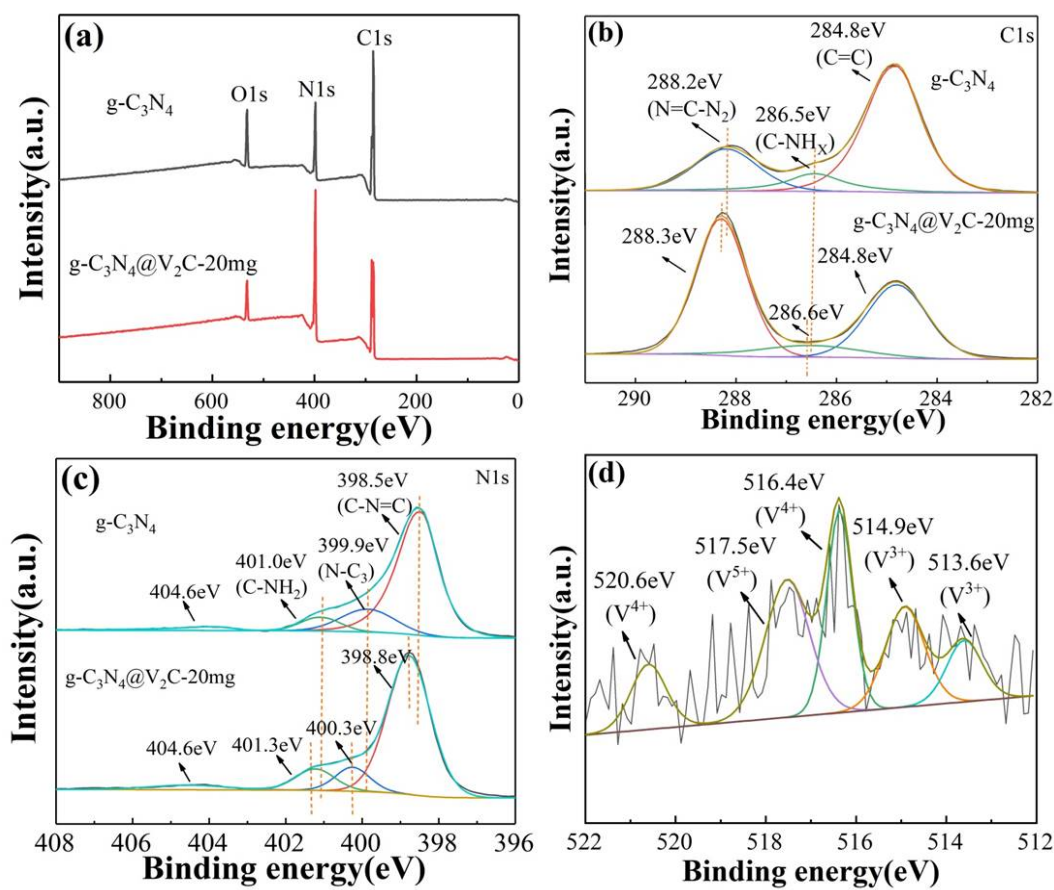

**Fig. 5.** (a) XPS spectra, (b) C1s high-resolution spectra, and (c) N1 high-resolution spectra of g-$C_3N_4$ and g-$C_3N_4$@$V_2$C-20 mg. (d) V2p high-resolution XPS spectrum of g-$C_3N_4$@$V_2$C-20 mg.

3.2 Bandgap and optoelectronic characterization

Fig. 6a presents the UV–Vis absorption spectra of $V_2$C MXene, g-$C_3N_4$, and g-$C_3N_4$@$V_2$C-20 mg. The absorption edges of g-$C_3N_4$@$V_2$C-20 mg and pure g-$C_3N_4$ are clearly visible at 440 nm. The composite exhibits improved absorption properties in ultraviolet and visible regions, which was attributed to the broadband absorption of $V_2$C MXene across the solar spectrum. Moreover, the absorbance of g-$C_3N_4$@$V_2$C-20 mg is approximately two times higher than that of g-$C_3N_4$ under visible light (450–800 nm). Obtained through the Tauc-function bandgap simulation, the bandgaps of g-$C_3N_4$ and g-$C_3N_4$@$V_2$C-20 mg are 2.84 and 2.82 eV, respectively (Fig. 6b) [34]. The g-$C_3N_4$@$V_2$C-20 mg composite can absorb a greater range of wavelengths in the visible spectrum because of the narrower bandgap, suggesting that it may enhance the photocatalytic degradation efficiency of MO. Zhang et al. reported that $V_2$C MXene with –F and –OH end groups is a narrow-bandgap semiconductor. Thus, we determined that the bandgap of $V_2$C MXene is 0.42 eV (Fig. 6c) [35].

Fig. 6d presents the electrochemical impedance spectra of different photocatalysts. Overall, the smaller the arc radius, the smaller the impedance of charge transfer and the higher the charge transfer rate of the photocatalyst. Evidently, g-$C_3N_4$@$V_2$C-20 mg exhibits a lower arc radius than g-$C_3N_4$. The steady-state PL spectra and transient photocurrent response of g-$C_3N_4$@$V_2$C-20 mg and g-$C_3N_4$ were further measured to investigate the electron–hole pair separation efficiency and transport efficiency of the composite. As shown in Fig. 6e, under 350-nm excitation, both g-$C_3N_4$@$V_2$C MXene photocatalyst and g-$C_3N_4$ exhibit an emission peak at 470 nm. Notably, the peak intensity of g-$C_3N_4$@$V_2$C-20 mg is lower, suggesting that the electron–hole pair recombination frequency of the g-$C_3N_4$@$V_2$C-20 mg photocatalyst is lower and the number of photogenerated carriers is higher than those of g-$C_3N_4$. Additionally, after the addition of $V_2$C MXene, the photocurrent of the composites substantially increases, implying that g-$C_3N_4$@$V_2$C-20 mg exhibits considerably better photogenerated carrier separation capacity (Fig. 6f) [36].

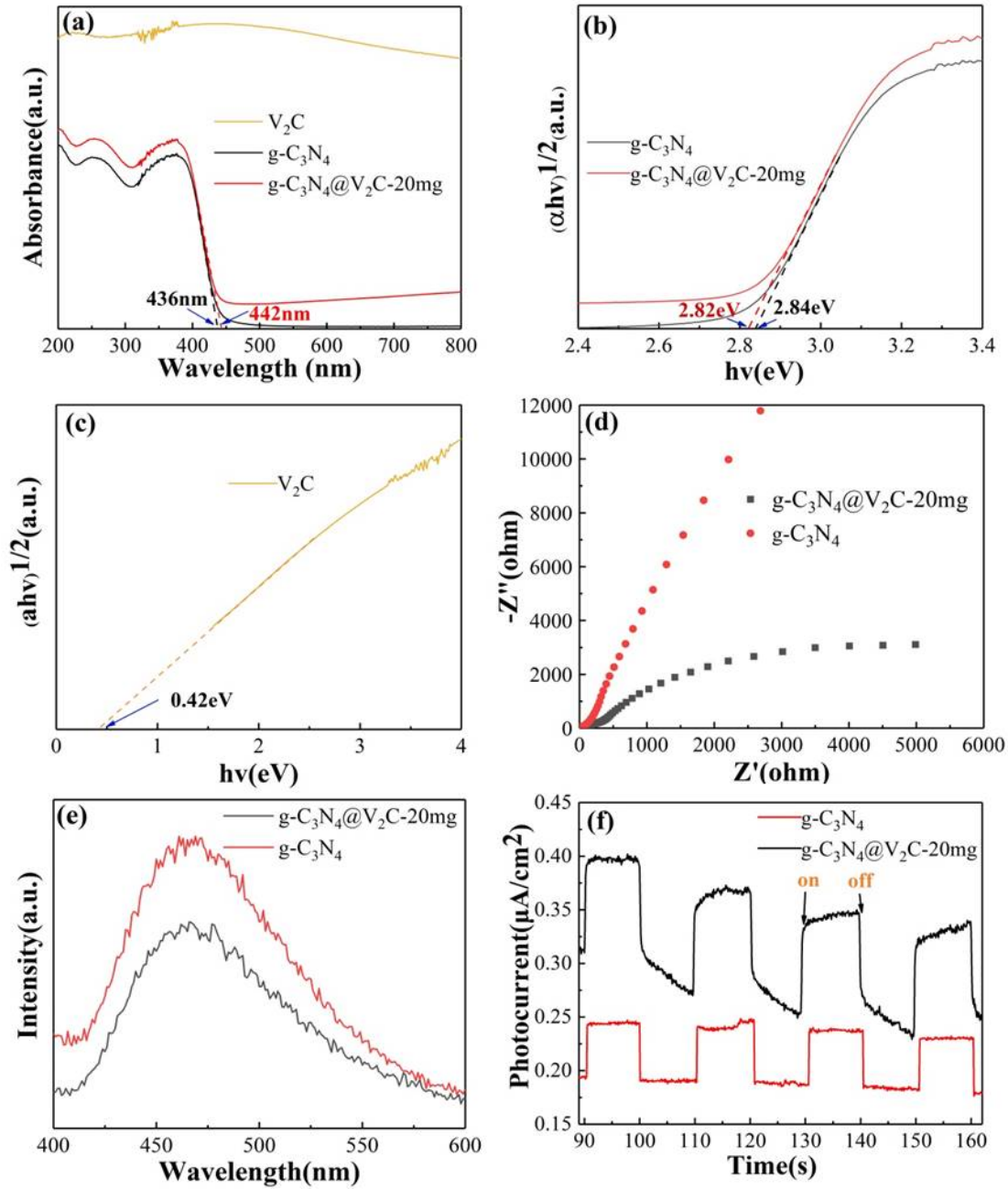

**Fig. 6.** (a–c) UV–Vis absorption spectra and Tauc-function bandgap simulations of g-C$_3$N$_4$, V$_2$C MXene, and g-C$_3$N$_4$@V$_2$C-20 mg. (d–f) Electrochemical impedance spectra, steady-state PL spectra, and photocurrent responses of g-C$_3$N$_4$ and g-C$_3$N$_4$@V$_2$C-20 mg.

3.3 Photocatalytic performance

The photocatalytic performances of V$_2$C MXene, g-C$_3$N$_4$, and g-C$_3$N$_4$@V$_2$C-X were investigated for different MO concentrations using a 300 W xenon lamp (Fig. 7a). The degradation efficiency was defined as $\eta = (C_0 - C) / C_0$, where $C_0$ represents the initial MO concentration and $C$ represents the MO concentration at time $t$. For each photocatalyst, three sets of parallel experiments were performed, and the data from the three sets of experiments were acquired and averaged, and this average value was utilized for graphing. The error value is the difference between the three values and the average. As shown in Fig. 7a, the concentration of MO negligibly changes when using pure $V_2C$ MXene. The degradation efficiency of MO by $g-C_3N_4$ is 60.4% after irradiation for 120 min. The MO degradation efficiency of $g-C_3N_4@V_2C$-10 mg, $g-C_3N_4@V_2C$-20 mg, and $g-C_3N_4@V_2C$-30 mg are 86.6%, 94.5%, and 88.8%, respectively, after irradiation for 120 min. The degradation efficiency of the $g-C_3N_4@V_2C$ MXene photocatalyst is 34.1% higher than that of $g-C_3N_4$. The contact between $V_2C$ MXene and $g-C_3N_4$ at a low amount of $V_2C$ MXene is not sufficient to produce a substantial number of exposed active centers and decrease the carrier separation efficiency. At the same time, an excessive amount of $V_2C$ MXene absorbs an excessive amount of visible light, constraining the effective light contact area of $g-C_3N_4$ and preventing the generation of a sufficient amount of photogenerated carriers in $g-C_3N_4$ to maintain an effective contact with dye molecules, thus reducing the photocatalytic efficiency. Therefore, the introduction of an appropriate amount of $V_2C$ MXene results in a high photocatalytic degradation efficiency of the photocatalyst.

The kinetic behavior of the degradation reaction may be expressed using the first-order kinetic equation $-\ln(C/C_0) = kt$, where $k$ and $t$ are the rate constant of the photocatalytic degradation process and time of illumination, respectively. The $k$ values for $V_2C$, $g-C_3N_4$, $g-C_3N_4@V_2C$-10 mg, $g-C_3N_4@V_2C$-20 mg, and $g-C_3N_4@V_2C$-30 mg were 0.0012, 0.00756, 0.01659, 0.02475, and 0.01936 $min^{-1}$, respectively. The $g-C_3N_4@V_2C$-20 mg photocatalyst shows the highest $k$ of $0.02475 min^{-1}$, which is 3.3 times higher than that of $g-C_3N_4$. At the same time, the $k$ values of $g-C_3N_4@V_2C$-10 mg and $g-C_3N_4@V_2C$-30 mg are 2.2 and 2.6 times, respectively, higher than that of $g-C_3N_4$. These results further confirm that $V_2C$ MXene is a promising cocatalyst for the photocatalytic degradation of dyes.

In addition to good photocatalytic degradation performance, a promising photocatalyst must also exhibit a certain degree of stability. After three repeated cycles (Fig. 7c), the photocatalytic degradation performance of MO by the $g-C_3N_4@V_2C$-20 mg catalyst did not considerably change and remained above 85%. Thus, the $g-C_3N_4@V_2C$ MXene photocatalyst has excellent photocatalytic performance and cycling stability. Moreover, as compared to other $g-C_3N_4$-based photocatalysts, the $g-C_3N_4@V_2C$ MXene photocatalyst in this work also had greater photocatalytic ability. The $V_2C$ MXene material is a potential co-catalyst for the photocatalytic degradation of dyes, which can be further substantiated based on this information.

To further investigate the mechanism through which the $g-C_3N_4@V_2C$ MXene photocatalyst promotes the degradation of pollutants (MO), radical trapping tests were performed on the $g-C_3N_4@V_2C$ MXene photocatalyst. The following compounds were used as scavengers of superoxide radicals ($\cdot O_2^-$), holes ($h^+$), electrons ($e^-$), and hydroxyl radicals ($\cdot OH$): benzoquinone (BQ), potassium iodide (KI), potassium dichromate ($Kr_2Cr_2O_7$), and isopropanol (IPA), respectively [37]. The MO without added free radical scavengers was almost completely degraded. As shown in Fig. 7d, the $C/C_0$ without the addition of radical scavengers is 0.055. After the addition of BQ, the photocatalytic performance decreased by approximately 80%, thus demonstrating that $·O_2^-$ is the primary active group in the degradation process. The $C/C_0$ after the addition of four other free radical scavengers—BQ, KI, $Kr_2Cr_2O_7$, and IPA—are 0.9, 0.42, 0.35, and 0.29, respectively. Therefore, the addition of KI, $Kr_2Cr_2O_7$, and IPA also inhibited the photocatalytic degradation of MO by the composites, indicating that $h^+$, $e^-$, and ·OH are also active groups in the photocatalytic degradation. Based on the above results, the degree of action of the active groups is $·O_2^- > h^+ > e^- > ·OH$.

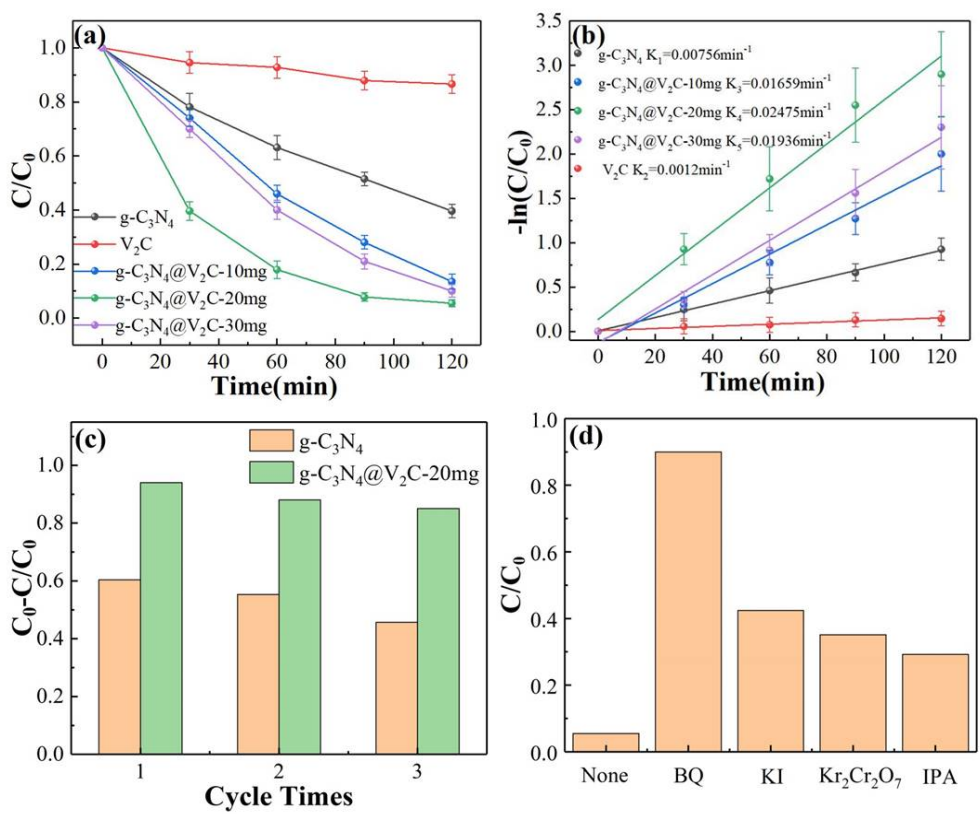

**Fig. 7.** (a) Photocatalytic degradation by $g-C_3N_4$, $V_2C$ MXene, $g-C_3N_4@V_2C$-10 mg, $g-C_3N_4@V_2C$-20 mg, and $g-C_3N_4@V_2C$-30 mg under visible light ($C$ and $C_0$ are MO concentrations after and before irradiation, respectively); (b) degradation kinetics of $g-C_3N_4$, $V_2C$ MXene, $g-C_3N_4@V_2C$-10 mg, $g-C_3N_4@V_2C$-20 mg, and $g-C_3N_4@V_2C$-30 mg under visible light; (c) cycling stability of $g-C_3N_4$ and $g-C_3N_4@V_2C$-20 mg in MO degradation under visible light; and (d) effect of different radical scavengers on the photocatalytic degradation by $g-C_3N_4@V_2C$-20 mg.

Table 1. Comparison of photocatalytic degradation efficiency of MO with different $g-C_3N_4$-based photocatalysts.

| Photocatalysts | Light | Degradation rate (%) | Time (min) | Pollutants | Reference |
| --- | --- | --- | --- | --- | --- |
| g-C$_3$N$_4$@V$_2$C-20mg | Vis | 94.5 | 120 | MO | This |
| BN/g-C$_3$N$_4$ | Vis | 80 | 90 | MO | [38] |
| g-C$_3$N$_4$/Ag/P$_3$HT | Vis | 75 | 480 | MO | [39] |
| Fe(III)-C$_3$N$_4$ | Vis | 80 | 70 | MO | [40] |
| GCN-3/VL/PDS | Vis | 89.6 | 40 | MO | [41] |
| ZnO HC@Ti$_3$C$_2$ | Vis | 87.8 | 60 | MO | [42] |
| MF-MXene/ppy | Vis | 91.75 | 240 | MO | [43] |

3.4 Possible photocatalytic mechanisms

The Mott–Schottky curve can be used for obtaining the flat-band potential ($E_{fb}$) of semiconductors. The intercept of the straight line with the horizontal axis is $E_{fb}$, which is considered the Fermi energy level. If the Fermi energy level is near the edge of the energy band, the obtained $E_{fb}$ can be considered an extreme value of the conduction band ($E_{CB}$, n-type) or valence band ($E_{VB}$, p-type) [44]. As shown in Fig. 8, g-C$_3$N$_4$, V$_2$C MXene, and g-C$_3$N$_4$@V$_2$C-20 mg exhibit positive slopes and are n-type semiconductors. With SCE as the reference electrode, the intersections of the approximated straight line with the horizontal axis for g-C$_3$N$_4$, V$_2$C MXene, and g-C$_3$N$_4$@V$_2$C-20 mg are −1.50, −1.29, and −1.36 eV, respectively. The SCE deviation is 0.24 eV compared to the standard hydrogen electrode. Thus, the $E_{CB}$ of g-C$_3$N$_4$, V$_2$C MXene, and g-C$_3$N$_4$@V$_2$C-20 mg have the following values: −1.26, −1.05, and −1.12 eV, respectively. Using empirical Equation (1), the $E_{CB}$ and $E_{VB}$ of a semiconductor can be obtained [45]. The determined $E_{VB}$ of g-C$_3$N$_4$, V$_2$C MXene, and g-C$_3$N$_4$@V$_2$C-20 mg are 1.58, −0.63, and 1.70 eV, respectively.

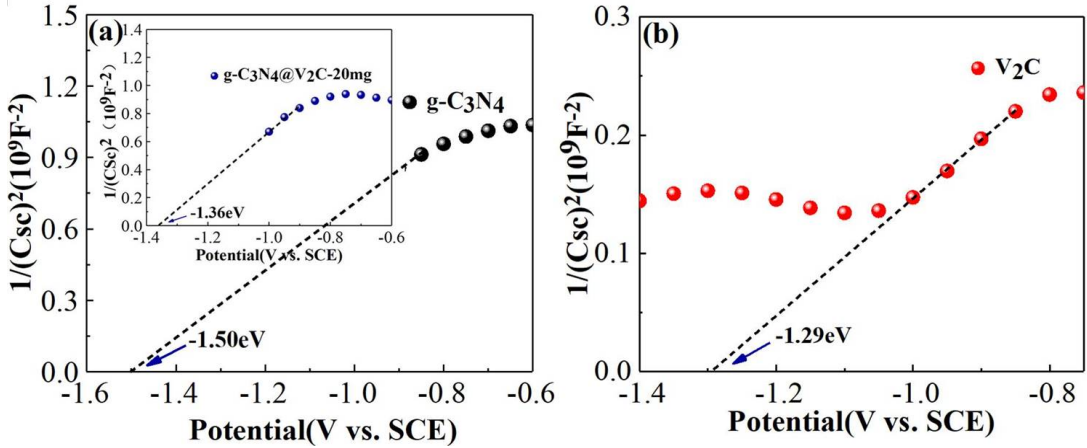

**Fig. 8.** Mott–Schottky curves for (a) g-C$_3$N$_4$ (the inset corresponds to g-C$_3$N$_4$@V$_2$C-20 mg) and (b) V$_2$C MXene.

$$E_{VB} = E_g + E_{CB} \tag{1}$$

$$2\,g-C_3N_4 + h\nu \rightarrow g-C_3N_4\,(e^-) + g-C_3N_4\,(h^+) \tag{2}$$
$$g-C_3N_4\,(e^-) + V_2C \rightarrow g-C_3N_4 + V_2C\,(e^-) \tag{3}$$

$$O_2 + V_2C\ (e^-) \rightarrow \cdot O_2^- + V_2C \tag{4}$$
$$\cdot O_2^- + H^+ \rightarrow \cdot HO_2^- \tag{5}$$
$$g - C_3N_4\ (h^+) + H_2O + \cdot O_2^- \rightarrow g - C_3N_4 + \cdot OH + h^+ \tag{6}$$
$$g - C_3N_4\ (h^+) + Oh^- \rightarrow \cdot OH \tag{7}$$
$$MO\ (dye) + \cdot OH^-/\cdot O_2^-/\cdot HO_2^- \rightarrow CO_2 + H_2O \tag{8}$$

Based on the theoretical analysis and characterization tests, a rational photocatalytic mechanism was suggested to illustrate the photocatalytic process of g-$C_3N_4$@$V_2C$ MXene photocatalyst on MO. Empirical equations (2)–(8) describe the conversion and formation of different active groups of g-$C_3N_4$@$V_2C$ MXene photocatalyst throughout photocatalytic degradation [46]. Fig. 9 illustrates how g-$C_3N_4$ generates photocarriers after absorbing visible light. Electrons ($e^-$) are transferred from the valence band ($E_{VB}$) to conduction band ($E_{CB}$), but the photocatalytic efficiency is hampered by quick recombination of a substantial number of electron–hole pairs at this time. When $V_2C$ MXene and g-$C_3N_4$ are tightly connected, $V_2C$ MXene acts as an electron acceptor at the better band edge position and acts as a cocatalyst in the photocatalytic degradation of pollutants (MO). Therefore, less recombination occurs between the photogenerated carriers ($e^-$–$h^+$), and a substantial number of electrons ($e^-$) rapidly migrate from the g-$C_3N_4$ conduction band ($E_{CB}$) to $V_2C$ MXene. These photogenerated electrons that have migrated could subsequently interact with organic molecules to produce $CO_2$ and ·OH. In addition, the band edge of MXene is lower than the band edge of ·$O_2^-$ (−0.33 eV), facilitating the excitation and production of superoxide radicals (·$O_2^-$). The valence band ($E_{VB}$) of g-$C_3N_4$ is higher than that of ·OH/$OH^-$ (2.29 eV). Thus, some of the generated ·$O_2^-$ react with $H^+$ to produce ·OH/$OH^-$ [47]. Finally, these active groups (·OH/·$O_2^-$) bind to MO molecules and degrade them to produce $CO_2$.

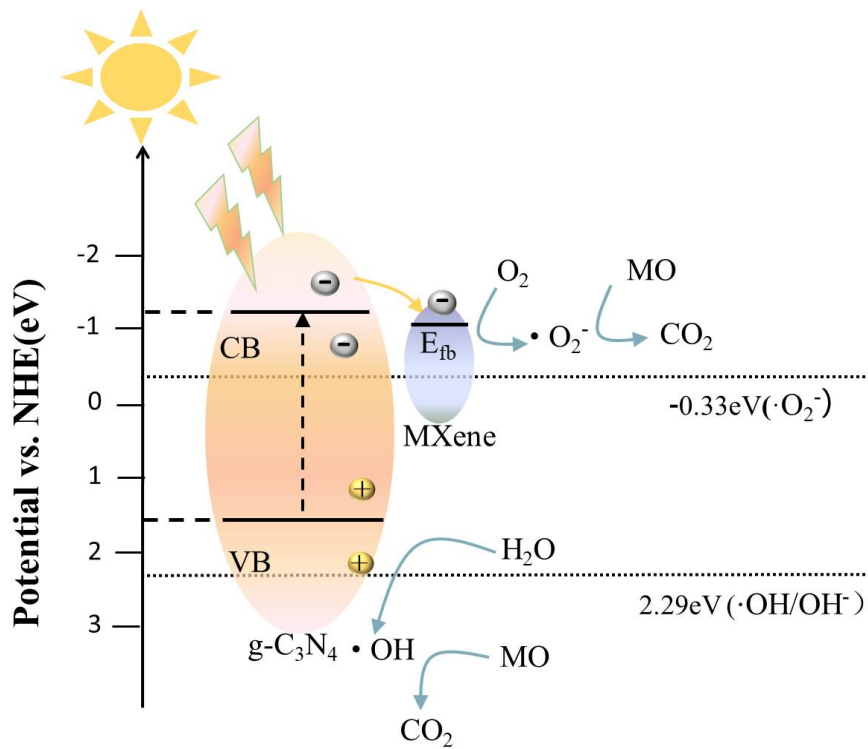

**Fig. 9.** Degradation mechanism by g-C$_3$N$_4$@V$_2$C MXene photocatalyst under visible light.

## 4 Conclusions

Herein, a novel photocatalyst with a closely connected interface, g-C$_3$N$_4$@V$_2$C MXene photocatalyst, was synthesized through modification of g-C$_3$N$_4$ with V$_2$C MXene. The degradation efficiency of MO by the g-C$_3$N$_4$@V$_2$C-20 mg composite is up to 94.5% after 120 min of exposure to visible light. The g-C$_3$N$_4$@V$_2$C MXene photocatalyst can swiftly transfer photogenerated electrons (e$^-$) from g-C$_3$N$_4$ to V$_2$C MXene, restrain the rapid recombination between electrons and holes (e$^-$–h$^+$), cause participation of a substantial number of photogenerated carriers in subsequent oxidation–reduction, and promote the photocatalytic degradation of MO. Moreover, the cycling experiment demonstrated the outstanding photocatalytic activity and cycling stability of g-C$_3$N$_4$@V$_2$C MXene photocatalyst. This work confirms that V$_2$C MXene is an excellent cocatalyst for degrading dyes in water and provides a new perspective for designing g-C$_3$N$_4$-based photocatalysts.

## CRediT authorship contribution statement

**Ruizheng Xu:** Investigation, Formal analysis, Writing – original draft. **Guiyu Wei：** Formal analysis. **Zhemin Xie：** Data curation. **Sijie Diao：** Investigation. **Jianfeng Wen:** Conceptualization, Funding acquisition. **Tao Tang:** Funding acquisition, Resources. **Li Jiang:** Supervision. **Ming Li:** Supervision, Writing – review & editing. **Guanghui Hu:** Writing – review & editing.

## Declaration of Competing Interest

The authors declare that they have no known competing financial interests or personal relationships that could have appeared to influence the work reported in this paper.

## Declaration of generative AI and AI-assisted technologies in the writing process

All authors of this thesis did not use generative artificial intelligence (AI) or AI-assisted technologies in the writing process.

## Acknowledgments

This work was financially supported by the National Natural Science Foundation of China (12164013), the Natural Science Foundation of Guangxi Province (2020GXNSFBA297125), the Science and Technology Base and Talent Special Project of Guangxi Province (AD21220029), Research Foundation of Guilin University of Technology (GUTQDJJ2019011).

materials, Chinese J. Catal. 42 (2021) 710-730.